\documentclass[11pt]{article}
\usepackage{epsfig}

\begin{document}
\begin{titlepage}

\hfill {\today}

\begin{center}
\ \\
{\Large \bf Skyrmions at finite density and temperature: the chiral phase transition }
\\
\vspace{.30cm}

Byung-Yoon Park$^{(a)}$, Hee-Jung Lee$^{(b)}$, and Vicente
Vento$^{(c)}$

\vskip 0.20cm
{(a) \it Department of Physics,
Chungnam National University, Daejon 305-764, Korea}\\
({\small E-mail: bypark@cnu.ac.kr})

{(b) \it School of Science Education,
Chungbuk National University, Cheongju Chungbuk 361-763, Korea}\\
({\small E-mail: hjl@chungbuk.ac.kr})

{(c) \it Departament de Fisica Te\`orica and Institut de F\'{\i}sica
Corpuscular}\\
{\it Universitat de Val\`encia and Consejo Superior
de Investigaciones Cient\'{\i}ficas}\\
{\it E-46100 Burjassot (Val\`encia), Spain} \\ ({\small E-mail:
Vicente.Vento@uv.es})

\end{center}
\vskip 0.3cm

\centerline{\bf Abstract}
The Skyrme model, an effective low energy theory rooted in large $N_c$ QCD,
has been applied to the study of dense matter.  Matter is described by
various crystal structures of skyrmions. When this system is heated, the dominating thermal
degrees of freedom are  the fluctuating pions. Taking  these mechanisms
jointly produces a description of
the chiral phase transition leading to the conventional phase diagram with
critical temperatures and densities in agreement with expected values.

\vskip 0.5cm

\vskip 0.3cm \leftline{Pacs: 11.30.Rd, 12.39.Dc, 14.40.Aq, 25.75.Nq}
\leftline{Keywords: skyrmion, dense matter, finite temperature, fluctuating pions, chiral phase transition}

\end{titlepage}

An important issue today is to understand the properties of
hadronic matter under extreme conditions, e.g., at high temperature
as in relativistic heavy-ion physics and/or at high density as in
compact stars. The phase diagram of hadronic matter turns out
richer than what has been predicted by
perturbative Quantum Chromodynamics (QCD) \cite{QCDphasetransition}.
Lattice QCD, the main computational tool accessible to highly
nonperturbative QCD, has provided much information on the
the finite temperature transition, such as  the value of the critical temperature, the type of
equation of state,  etc\cite{Karsch07}. However, due to a notorious `sign problem',
lattice QCD could not be applied to study dense
matter. Only in the last few years, it has become possible to
simulate QCD with small baryon density \cite{Fromm:2008ab}.
Chiral symmetry is a flavor symmetry of QCD which plays an essential
role in hadronic physics. At low temperatures and densities it is spontaneously
 broken leading to the existence of the pion.  Lattice studies seem to imply that chiral
 symmetry is restored in the high temperature and/or high baryon density phases and
 that it may go hand-in-hand with the confinement/deconfinement transition.
The quark condensate $\langle \bar{q} q\rangle$ of QCD is an order parameter
of this symmetry and decreases to zero when the symmetry is
restored.

The Skyrme model, an effective low energy theory rooted in large $N_c$ QCD,
has been applied to the dense matter studies
\cite{LPMRV02,PRV03,PRV08,LPRV03,LPRV04,KP05,KCP05}.
The model does not have explicit quark and gluon degrees of freedom,
and therefore one can not investigate the
confinement/deconfinement transition directly, but we may study
 the chiral symmetry restoration transition which occurs close by.
 The main ingredient associated with chiral symmetry is the pion, the Goldstone
boson associated with the spontaneously broken phase. The
various patterns in which the symmetry is realized in QCD will be
directly reflected in the in-medium properties of the pion and
consequently in the properties of the skyrmions made of it.

The classical nature of skyrmions enables us to construct the dense
system quite conveniently by putting more and more skyrmions into a
given volume. Then, skyrmions shape and arrange themselves to
minimize the energy of the system. The ground state configuration of
skyrmion matter are crystals. At low density it is made of
well-localized single skyrmions \cite{Klebanov85}. At a critical
density, the system undergoes a structural phase transition to a new
kind of crystal. It is made of `half-skyrmions' which are still
well-localized but carry only  half winding number. In the
half-skyrmion phase, the system develops an additional symmetry
which leads to a vanishing average value for $\langle \sigma
\rangle$ \cite{Goldhaber87}. In the studies of the late
80's \cite{Jackson88}, the vanishing of $\langle \sigma \rangle$ was
interpreted as chiral symmetry restoration by assuming that $\langle
\sigma \rangle$ is related to the QCD order parameter $\langle
\bar{q}q\rangle$. However, in Ref.~\cite{LPMRV02}, it was shown that the
vanishing of $\langle \sigma \rangle$ cannot be an indication of a
genuine chiral symmetry restoration, because the decay constant of
the pion fluctuating in such a half-skyrmion matter does not vanish.
This was interpreted as a signal of the appearance of  a pseudo-gap phase
similar to what happens in high $T_c$ superconductors \cite{Franz02}.

The puzzle was solved in Ref.~\cite{LPRV03} by incorporating a
suitable degree of freedom, the dilaton field $\chi$, associated
to the scale anomaly of QCD. The dilaton field takes
over the role of the order parameter for  chiral symmetry
restoration. As the density of skyrmion matter increases, both
$\langle \sigma\rangle$ and $\langle \chi \rangle$ vanish (not
necessarily at the same critical density). The effective decay
constant of the pion fluctuation vanishes only when $\langle
\chi\rangle$ becomes zero. It is thus the dilaton field which
provides the mechanism for chiral symmetry restoration.

Contrary to lattice QCD, there are few studies on the
temperature dependence of skyrmion matter.
Skyrmion matter has been heated up to melt the crystal into a
liquid to investigate the crystal-liquid phase transition
\cite{Kaelbermann00,Schwindt:2002we} a phenomenon which is irrelevant to the restoration
of chiral symmetry. In here, we study skyrmion matter at {\em finite density
and temperature}. We obtain the phase diagram
describing the realization of the chiral symmetry.

The basic strategy of our approach is as follows. We describe dense skyrmion matter as a certain
crystal structure. Thereafter, we let the temperature rise. The dominating thermal
degrees of freedom in this process are the fluctuating pions which contribute a
thermal energy proportional to $T^4$. We neglect the thermal fluctuations of
the heavy skymions (translations, rotations and vibrations) which are proportional to $T$.
We show that this strategy leads to a familiar phase diagram with
critical temperature and density in agreement with the expected values.

The simplest model Lagrangian containing the
inevitable ingredients to make the calculation respect the flavor properties of $QCD$ reads \cite{EL85,BR91}
\begin{equation}
{\cal L} = \chi^2 {\cal L}_\sigma +{\cal L}_{\mbox{\scriptsize sk}}
- {\cal V}(\chi),
\label{L}\end{equation}
with
$$ {\cal L}_\sigma
= \frac{f^2_\pi}{4} \mbox{tr} (\partial^\mu U^\dagger\partial_\mu U),
\eqno(\ref{L}\mbox{{\rm a}})$$
$$ {\cal L}_{\rm\scriptstyle sk} =
\frac{1}{32e^2_{\mbox{\scriptsize sk}}} \mbox{tr} [U^\dagger
\partial_\mu U, U^\dagger \partial_\nu ]^2
\eqno(\ref{L}\mbox{\rm b})$$
$${\cal V}(\chi) =
\frac{m_\chi^2 f_\chi^2}{4} \textstyle \left\{ \chi^4 \left( \ln
\chi -  \frac14 \right) + \frac14 \right\}.
\eqno(\ref{L}\mbox{\rm c})$$
In this equation, $U=\exp(i\vec{\tau}\cdot\vec{\pi}/f_\pi)(\in SU(2))$ is
a nonlinear realization of the pion fields and $\chi$ represents the
dilaton field. The model Lagrangian contains
four parameters, $f_\pi$, $e_{\mbox{\scriptsize sk}}$, $m_\chi$ and
$f_\chi$, i.e. the pion decay constant, the Skyrme parameter, the dilaton decay
constant and the dilaton mass, respectively. We choose in what follows $f_\pi = 93$ MeV,
$e_{\mbox{\scriptsize sk}}=4.75$, $m_\chi=720$ MeV and $f_\chi=210$
MeV (a discussion on the values of these parameters can be found in Ref. \cite{LPRV03}).
For simplicity we neglect the pion mass and take the  $\chi$ field as a constant in space and
time. Still, the `constant' dilaton field controls the chiral phase
transition through the potential energy term. In the vacuum, this term has its
minimum at $\chi=1$. At finite density and temperature, skyrmions
and thermal pions contribute to the term proportional to $\chi^2$
and the system undergoes a {\em dynamical} phase transition when
this term is larger than the potential energy.  While the specific values for the
critical temperature and density  depend strongly on the chosen parameters,
the phase transition scenario remains qualitatively the same for all reasonable values.

Skyrmion matter is described by the energy
of a single skyrmion:
\begin{equation}
{\cal E}(\rho) = \chi^2 {\cal E}_\sigma  + {\cal E}_{\mbox{\scriptsize sk}}
 + {\cal V}(\chi ) V,
\label{E}\end{equation}
where
$$ {\cal E}_\sigma(\rho) = \int_V d^3 r\ \mbox{tr}
(\partial_i U_0 \partial_i U^\dagger_0),
\eqno(\ref{E}\mbox{a})$$
$$ {\cal E}_{\mbox{\scriptsize sk}}(\rho) = \int_V d^3 r\ \mbox{tr}
[U^\dagger_0 \partial_i U_0, U^\dagger_0 \partial_j U_0 ]^2,
\eqno(\ref{E}\mbox{b})$$
$\rho$ is the baryon number density of
the system and $V$ is the volume occupied by a single skyrmion; i.e., $\rho = 1/V$.
$U_0(\vec{r})$ is the static skyrmion
configuration that minimizes the system energy ${\cal E}$ for a given $\rho$.
Note that both the potential energy
terms, which contains an explicit $1/\rho(=V)$ factor, and ${\cal
E}_{\sigma,\mbox{\scriptsize sk}}$, through $U_0(\vec{r})$, depend on the baryon
number density.

What happens if we  heat up the system? Naively, as the temperature increases, the kinetic
energy of the skyrmions increases and the skyrmion crystal
begins to melt. The kinetic
energy associated with the translations, vibrations and
rotations of the skyrmions is proportional to $T$. This mechanism
leads to a solid-liquid-gas phase transition of the skyrmion system.
However, we are not interested in this transition but in the chiral
symmetry restoration transition. A new mechanism must be incorporated
to describe it and it happens to be the thermal excitation of the pions in the medium.
This phenomenon is proportional to $T^4$ and therefore
dominates the absorption of heat. We only consider the latter in our discussion.

The pressure of the non-interacting pions is given by \cite{BK95}
\begin{equation}
P=\frac{\pi^2}{30} T^4,
\end{equation}
where we have taken into account the contributions from three pions,
$\pi^+, \pi^0, \pi^-$.
This term contributes to the energy per single skyrmion volume as $3PV \chi^2$.
The factor 3 comes from the fact that our pions are massless.
The kinetic energy of the pions arises from ${\cal L}_\sigma$, and therefore scaling
implies that it should carry a factor $\chi^2$. After including the
thermal pions, the energy lodged around a single
skyrmion can be expressed as
\begin{equation}
{\cal E}(\rho, T) =
\left({\cal E}_\sigma(\rho) + \frac{\pi^2}{10}T^4 V \right) \chi^2
+ {\cal E}_{\mbox{\scriptsize sk}}(\rho)
+ {\cal V}(\chi )V.
\label{E1}\end{equation}
In order to complete our calculation we have to determine the values of $U_0(\vec{r})$ and $\chi$ that minimize
${\cal E}(\rho,T)$ for a given density and temperature.

Chiral restoration will occur when $\chi$ vanishes.
How does this take place in hot and dense matter? In eq.(\ref{E1}), the
potential energy of the dilaton is finite and its magnitude
decreases as the density $\rho$ increases ($V$ decreases). The potential energy has a
potential hill at $\chi =0$ and  a valley at $\chi = 1$. Beyond
$\chi = 1$, it increases monotonically.  Besides the potential term there is another term  proportional to
$\chi^2$. As the density increases the ${\cal E}_\sigma$ contribution becomes relatively important.
The thermal pion contribution is proportional
to $T^4/\rho$. Thus, the term proportional to $\chi^2$ increases to
overcome the potential hill.

Let us look into this phenomenon in detail. For a given $\rho$ and $T$, explicit dependence of eq.(\ref{E1}) on
$\chi$ can be presented as
\begin{equation}
{\cal E}  = \textstyle a \chi^2 + b + c (\chi^4(\ln \chi - \frac14)+\frac14).
\label{E_work}\end{equation}
One may easily match the coefficients $a$, $b$ and $c$ with eq.(\ref{E1}).
${\cal E}$ has two local minima at $\chi=0$ (chirality symmetric phase)
and at $\chi = {\chi}_0 (\neq 0)$ (chirality broken phase) satisfying
\begin{equation}
a + 2c {\chi}_0^2 \ln {\chi}_0 = 0.
\label{X0}\end{equation}
The energy difference
$\Delta {\cal E} \equiv {\cal E}({\chi=0}) -{\cal E}(\chi ={\chi}_0)$ is
obtained as
\begin{equation}
\Delta {\cal E}=  \textstyle \frac14 {\chi}_0^2 (2 a - c
{\chi}_0^2).
\end{equation}
The phase transition happens when $\Delta{\cal E} =0$; that is,
\begin{equation}
2 a - c {\chi}_0^2 = 0. \label{PT}\end{equation} Combining
eqs.(\ref{X0}) and (\ref{PT}), we get ${\chi}_0$ at the critical
point as
\begin{equation}
{\chi}_0^c = e^{-1/4}.
\end{equation}
Note that the phase transition happens suddenly from a non-vanishing
$\chi= {\chi}_0^c$ value to $\chi= 0$. It is therefore a first
order phase transition. However, this result is just a peculiarity of the used dilaton potential ${\cal V}(\chi)$.
If we had taken ${\cal V}(\chi) \sim (\chi^2-1)^2$,
the phase transition would have been second order.

Using the above equations at the critical point, we get
\begin{equation}
\rho^c {\cal E}_\sigma  + \frac{\pi^2}{10} T_c^4  = \frac{f_\chi^2
m_\chi^2}{8e^{1/2}} .
\label{PTRT}\end{equation}
which leads to
\begin{equation}
T_c =  \left( \frac{10}{\pi^2} \left(\frac{f_\chi^2
m_\chi^2}{8e^{1/2}}
 -\rho^c{\cal E}_\sigma (\rho^c)\right) \right)^{1/4}
\label{Tc}\end{equation}
For $\rho=0$ (zero density),
our estimate for the critical temperature is
\begin{equation}
T_c =  \left( \frac{10}{\pi^2} \frac{f_\chi^2 m_\chi^2}{8e^{1/2}}
\right)^{1/4} \sim  \mbox{205 MeV},
\end{equation}
where we have substituted $f_\chi=210$ MeV, $m_\chi=720$ MeV.
It is remarkable that this naive toy model leads to $T_c
\sim 200$ MeV, which is quite close to that obtained by lattice QCD \cite{Karsch07}
and in agreement with the data \cite{RHIC}.

\begin{figure}
\centerline{\epsfig{file=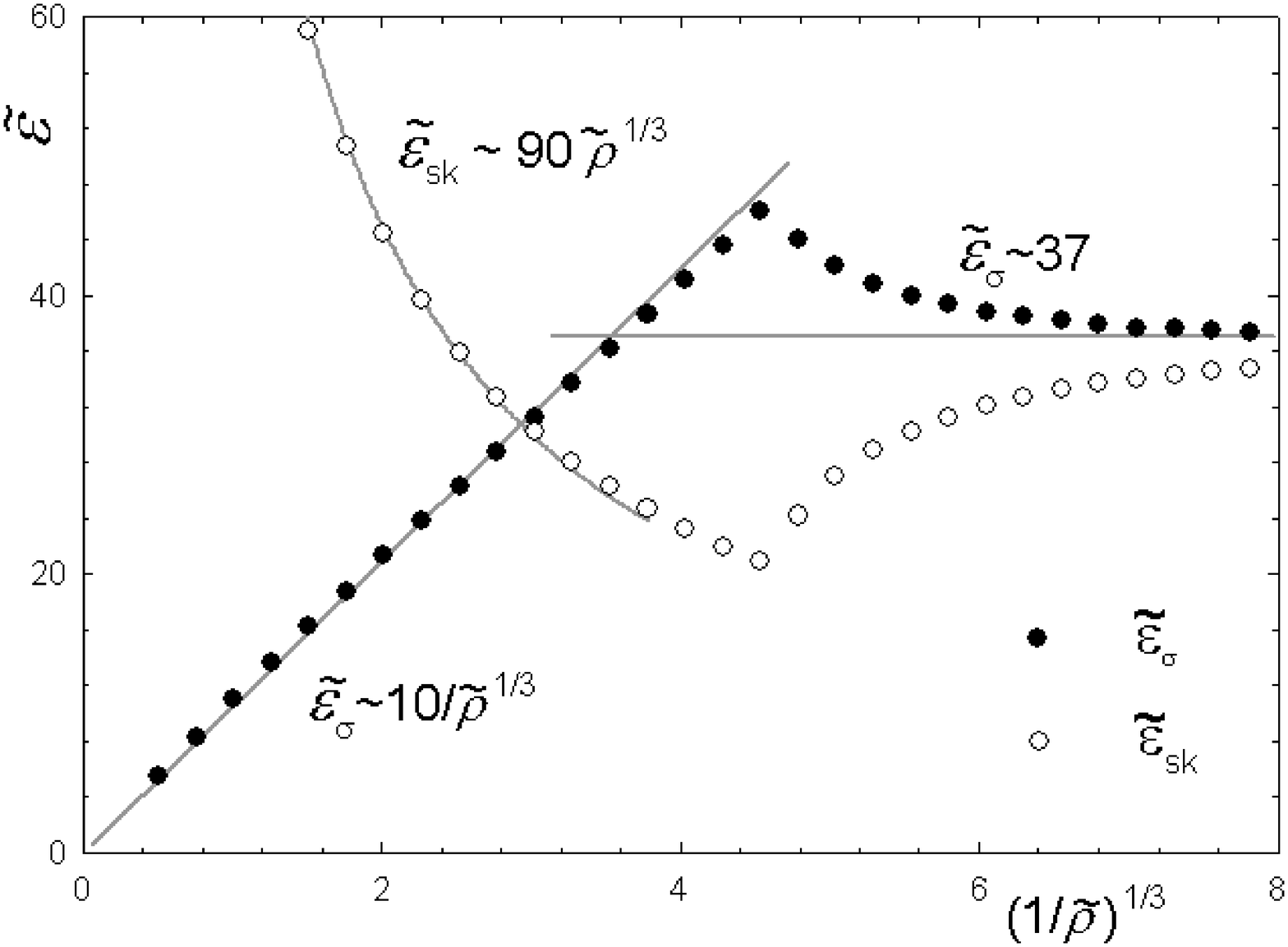,width=12cm,angle=0}}
\caption{$\tilde{\cal E}_\sigma$ and $\tilde{\cal E}_{\mbox{\scriptsize
sk}}$ as functions of $1/\tilde{\rho}^{1/3}$. }
\end{figure}

To complete our study  we compute ${\cal E}_\sigma$ as a function
of $\rho$ by minimizing numerically ${\cal E}(\rho, T)$  with respect to $U_0(\vec{r})$ and $\chi$.
An useful approximation may be obtained by fixing $\chi=1$ and minimizing
${\cal E}_\sigma(\rho) + {\cal E}_{\mbox{\scriptsize sk}}(\rho)$  with respect to $U_0(\vec{r})$. Since $\chi$ varies only in
the range $e^{-1/4}\sim 0.8 <\chi< 1$, before the system undergoes the phase transition, the approximation is quite reasonable.
These minimum energy configurations can be found by various
methods\cite{Kugler89}, and they turn out to be single skyrmion FCC (Face Centered Cubic) crystals at low density
and half-skyrmion CC (Centered Cubic) crystals at higher density.

\begin{figure}
\centerline{\epsfig{file=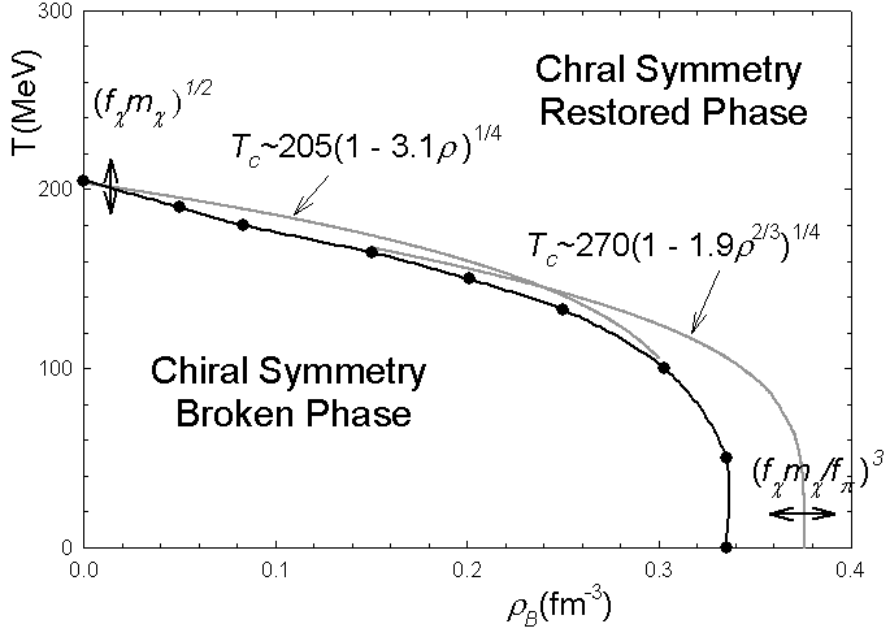,width=12cm,angle=0}}
\caption{The chiral phase transition. The solid line shows the exact calculation,
while the gray lines two approximate estimates. }
\end{figure}

The dependence of ${\cal E}_\sigma$
and ${\cal E}_{\mbox{\scriptsize sk}}$ on
$f_\pi$ and $e_{\mbox{\scriptsize sk}}$ can be separated out explicitly as
\begin{equation}
{\cal E}_{\sigma,\mbox{\scriptsize sk}}(\rho) =
(f_\pi/e_{\mbox{\scriptsize sk}}) \tilde{\cal E}_{\sigma,
\mbox{\scriptsize sk}}(\tilde{\rho}),
\end{equation}
where $\tilde{\rho} = \rho /(e_{\mbox{\scriptsize sk}}f_\pi)^3$ and
$\tilde{\cal E}_{\sigma, \mbox{\scriptsize sk}}$ are dimensionless
quantities which do not depend on
$e_{\mbox{\scriptsize sk}}$ and $f_\pi$.
Shown in Fig. 1 are the
numerical values on $\tilde{\cal E}_\sigma$ and $\tilde{\cal
E}_{\mbox{\scriptsize sk}}$. In order to show their dominant
asymptotic behaviors, we present various values of $\tilde{\cal E}$ as a function of
$1/(\tilde{\rho})^{1/3}$. For high densities $\tilde{\cal E}_\sigma$
is linear in $1/(\tilde{\rho})^{1/3}$, while for low densities it
stays constant, implying that at a low density the interaction between skyrmions is
negligible. We parameterize ${\cal E}_\sigma$ as
\begin{equation}
{\cal E}_\sigma = \left\{
\begin{array}{ll}
\displaystyle 10 f_\pi^2 /\rho^{1/3},  &
\rho > \rho_0 \\
\displaystyle 36 f_\pi / e_{\mbox{\scriptsize sk}} , &
\rho< \rho_0,
\end{array}
\right. \label{Asy}\end{equation} where
$\rho_0 =(e_{\mbox{\scriptsize sk}}f_\pi/3.6)^3$.

Using eq.~(\ref{Asy}) for ${\cal E}_\sigma$
and returning to  eq.~(\ref{PTRT}) we obtain the critical density for chiral
symmetry restoration at zero temperature as
\begin{equation}
\rho^c (T=0) = \left( \frac{f_\chi^2 m_\chi^2 }{8e^{1/2}}
\frac{1}{10f_\pi^2} \right)^{3/2} \sim 0.37 \mbox{ fm}^{-3}.
\label{Rho}
\end{equation}
Since $\rho_0 = 0.24$ fm$^{-3} < \rho^c(T=0) $ we have been consistent when using
the high density formula for $\tilde{\cal E}_\sigma$

The resulting critical density $\rho^c(T=0) \sim 0.37$ fm$^{-3}$ is
only twice  normal nuclear matter density and it is
low with respect to the expected values.  This result is not problematic since  $\rho^c (T=0) \sim (f_\chi m_\chi / f_\pi)^3$
and $T_c^{\rho=0} \sim (f_\chi m_\chi )^{1/2}$ and therefore the description of the phase transition does not depend on
the specific values of the parameters.  Moreover, small changes in them may  lead to
larger values for the critical density.

For a finite density smaller than $\rho^c (T=0)$, we obtain the
corresponding critical temperature by substituting the asymptotic
formulas (\ref{Asy}) for ${\cal E}_{\sigma}$,

\begin{eqnarray}
T_c & = & T_c(\rho=0)\;(1 - 3.09(fm^3)\; \rho_c)^{1/4} \;\;\; \mbox{      for  }  \rho < \rho_0 \\
T_c &= &T_c(\rho=0)\;(1 - 1.92(fm^2)\; \rho_c^{2/3})^{1/4}  \mbox{  for  }  \rho > \rho_0
\end{eqnarray}
The results of these two curves are  shown in Fig.~2 by  gray lines.

The exact calculation (black dots connected by black line in Fig.~2) has been obtained
numerically by minimizing the energy eq.(\ref{E1}).
The resulting phase diagram  has the same shape but the temperatures
and densities are generally smaller than in the approximate estimates shown in Fig. 2.

There is a long history of success of the Skyrme model in describing the chiral symmetry
restoration phase transition for dense hadronic matter \cite{LPMRV02,PRV03,PRV08,LPRV03,LPRV04,KP05,KCP05}.
The  model might also be a viable description of the new
phenomenology that is being proposed for dense/hot matter \cite{McLerran:2007qj}.
The time had come for an analysis of its behavior with temperature.
We have presented a first description of the chiral restoration phase
transition in the temperature-density plane, whose main ingredient
is that the dominant mechanism is the absorption of heat  by the fluctuating pions in the
background of crystal skyrmion matter. This description
leads to a phase transition whose dynamical structure  is parameter independent
and whose shape resembles much the conventional confinement/deconfiment phase transition.
Moreover, for parameter values close the conventional ones, we obtain
the expected critical temperatures and densities.
Further investigation in these
matters is required since it is becoming apparent that the path  to the phase transition is
not without new physics and that this `bottom up' approach might be useful to obtain
interesting new physics \cite{McLerran:2008ux}.

\section*{Acknowledgements}
Byung-Yoon Park was supported by a research fund from Chugnam
National University, Hee-Jung Lee by the research grant of
Chungbuk National University in 2008 and Vicente Vento by
grant FPA2007-65748-C02-01 from Ministerio de Ciencia e Innovaci\'on.

\end{document}